\documentclass[10pt,final,conference,leqno]{IEEEtran}
\usepackage[utf8]{inputenc}
\usepackage{amsmath}
\usepackage{amsthm}
\usepackage{amsfonts}
\usepackage{amssymb}

\usepackage{tikz}

\usepackage{mathtools}  % for mathllap

\usepackage{url}

\usepackage{verbatim}

\newtheorem{lemma}{Lemma}%[section]

\newtheorem{thm}[lemma]{Theorem}
\newtheorem*{thm*}{Theorem}

\theoremstyle{definition}

\theoremstyle{remark}

\newcommand{\TCI}{\widetilde{CI}}
\newcommand{\TSI}{\widetilde{SI}}
\newcommand{\TUI}{\widetilde{UI}}

\DeclareMathOperator{\AND}{AND}
\DeclareMathOperator{\XOR}{XOR}
\DeclareMathOperator{\bit}{bit}

\newcommand{\PIdrei}[9]{
    \node (X123) at (3,6)        {#1};
    \node (X12) at (1.5,5)       {#2};
    \node (X13) at (3  ,5)       {#3};
    \node (X23) at (4.5,5)       {#4};
    \node (X12X13) at (1.5,4)    {#5};
    \node (X12X23) at (3  ,4)    {#6};
    \node (X13X23) at (4.5,4)    {#7};
    \node (X1) at (0.5,3)        {#8};
    \node (X2) at (2. ,3)        {#9};
    \PIdreiTeilzwei
}
\newcommand{\PIdreiTeilzwei}[9]{
    \node (X3) at (3.5,3)        {#1};
    \node (X12X13X23) at (6 ,3)  {#2};
    \node (X1X23) at (1.5,2)     {#3};
    \node (X2X13) at (3  ,2)     {#4};
    \node (X3X12) at (4.5,2)     {#5};
    \node (X1X2) at (1.5,1)      {#6};
    \node (X1X3) at (3  ,1)      {#7};
    \node (X2X3) at (4.5,1)      {#8};
    \node (X1X2X3) at (3,0)      {#9};
    \begin{scope}[->,>=stealth]
      \path (X123) edge (X12);  \path (X123) edge (X13);  \path (X123) edge (X23);
      \path (X12) edge (X12X13);  \path (X12) edge (X12X23);
      \path (X13) edge (X12X13);  \path (X13) edge (X13X23);
      \path (X23) edge (X12X23);  \path (X23) edge (X13X23);
      \path (X12X13) edge (X1);  \path (X12X13) edge (X12X13X23);
      \path (X12X23) edge (X2);  \path (X12X23) edge (X12X13X23);
      \path (X13X23) edge (X3);  \path (X13X23) edge (X12X13X23);
      \path (X1) edge (X1X23);  \path (X12X13X23) edge (X1X23);
      \path (X2) edge (X2X13);  \path (X12X13X23) edge (X2X13);
      \path (X3) edge (X3X12);  \path (X12X13X23) edge (X3X12);
      \path (X1X23) edge (X1X2);  \path (X1X23) edge (X1X3);
      \path (X2X13) edge (X1X2);  \path (X2X13) edge (X2X3);
      \path (X3X12) edge (X1X3);  \path (X3X12) edge (X2X3);
      \path (X1X2) edge (X1X2X3);  \path (X1X3) edge (X1X2X3);  \path (X2X3) edge (X1X2X3);
    \end{scope}
}

\AtBeginDocument{%
 \abovedisplayskip=5pt plus 2pt minus 3pt
 \abovedisplayshortskip=0pt plus 3pt
 \belowdisplayskip=5pt plus 2pt minus 3pt
 \belowdisplayshortskip=0pt plus 3pt
}

\author{Johannes Rauh, Nils Bertschinger, Eckehard Olbrich, J{\"u}rgen Jost  \\
\small \{jrauh,bertschinger,olbrich,jjost\}@mis.mpg.de \\
\small Max Planck Institute for Mathematics in the Sciences, Leipzig, Germany}
\title{Reconsidering unique information: \\
  {% \huge
    Towards a multivariate information decomposition}}
\begin{document}
\maketitle

\begin{abstract}
  The information that two random variables $Y$, $Z$ contain about a
  third random variable~$X$ can have aspects of \emph{shared
    information} (contained in both $Y$ and $Z$), of
  \emph{complementary information} (only available from $(Y,Z)$
  together) and of \emph{unique information} (contained exclusively in
  either $Y$ or $Z$).  Here, we study measures $\TSI$ of shared,
  $\TUI$ unique and $\TCI$ complementary information introduced by
  Bertschinger \emph{et
    al.}~\cite{BROJA13:Quantifying_unique_information} which are % well
  motivated from a decision theoretic perspective. % In particular, 
  We find that in most cases the intuitive rule that \emph{more variables
    contain more information} applies, with the exception that $\TSI$
  and $\TCI$ information are not monotone in the target
  variable~$X$. Additionally, we show that it is not possible to
  extend the bivariate information decomposition into $\TSI$, $\TUI$ and $\TCI$ to a non-negative
  decomposition on the partial information lattice of Williams and
  Beer~\cite{WilliamsBeer:Nonneg_Decomposition_of_Multiinformation}. Nevertheless,
  the quantities $\TUI$, $\TSI$ and $\TCI$ have a well-defined
  interpretation, even in the multivariate setting.
\end{abstract}

\section{Introduction}

Consider three random variables $X,Y,Z$ with finite state spaces. % $\Xcal,\Ycal,\Zcal$.
Suppose that we are interested in the value of~$X$, but we can only observe $Y$ or~$Z$.  If the tuple $(Y,Z)$ is not
independent of~$X$, then the values of $Y$ or $Z$ or both of them contain information about~$X$.  The information about
$X$ contained in the tuple $(Y,Z)$ can be distributed in different ways.  For example, it may happen that $Y$ contains
information about~$X$, but $Z$ does not, or vice versa.  In this case, it would suffice to observe only one of the two
variables $Y,Z$, namely the one containing the information.  It may also happen, that both $Y$ and $Z$ contain different
information, so it would be worthwhile to observe both of the variables.  If both $Y$ and $Z$ contain the same
information about~$X$, we could chose to observe either $Y$ or~$Z$.  Finally, it is possible that neither $Y$ nor $Z$
taken for itself contains any information about~$X$, but together they contain information about~$X$.  This effect is
called \emph{synergy}, and it occurs, for example, if all variables $X,Y,Z$ are binary, and $X = Y\XOR Z$.  In general,
all effects may be present at the same time.  That is, the information that $(Y,Z)$ has about~$X$ is a combination of
\emph{shared information} $SI(X:Y;Z)$ (information contained both in $Y$ and in $Z$), \emph{unique information}
$UI(X:Y\setminus Z)$ and $UI(X:Z\setminus Y)$ (information that only one of $Y$ and $Z$ has) and
\emph{synergistic} or \emph{complementary information} $CI(X:Y;Z)$ (information that can only be retrieved when
considering $Y$ and $Z$ together)\footnote{It is often assumed that these three types of information are everything there is,
but one may ask, of course, whether there are further types of information.}.

Many people have tried to make these ideas precise and quantify the amount of unique information, shared information or
complementary information.  In particular, neuro-scientists have struggeled for a long time to come up with a suitable
measure of synergy;
see~\cite{SchneidmanBialekBerry03:Synergy_and_redundancy_in_population_codes,LathamNirenberg05:Synergy_and_redundancy_revisited}
and references therein.  A promising conceptual point of view was taken
in~\cite{WilliamsBeer:Nonneg_Decomposition_of_Multiinformation} by Williams and Beer, who developped the framework of the \emph{partial information lattice} to
define a decomposition of the mutual information into non-negative parts with a well-defined interpretation.  Their work
prompted a series of other papers trying to improve these
results~\cite{GriffithKoch13:Quantifying,HarderSalgePolani13:Bivariate_redundant_information,BROJ13:Shared_information}.
We recall the definition of the partial information lattice in Section~\ref{sec:PI-vs-Id}.

In this paper we build on the bivariate information decomposition defined
in~\cite{BROJA13:Quantifying_unique_information}, which is defined as follows: Let $\Delta$ be the set of all joint
distributions of $X$, $Y$ and~$Z$, and for fixed $P\in\Delta$ let $\Delta_{P}$ be the subset of $\Delta$ that consists
of all distributions $Q\in\Delta$ that have the same marginal distributions on the pairs $(X,Y)$ and $(X,Z)$,
i.e. $Q(X=x,Y=y)=P(X=x,Y=y)$ and $Q(X=x,Z=z)=P(X=x,Z=z)$ for all possible values $x,y,z$.  Then we define
\begin{align*}
  \TUI(X:Y\setminus Z) & = \min_{Q\in\Delta_{P}} MI_{Q}(X:Y|Z), \\
  \TUI(X:Z\setminus Y) & = \min_{Q\in\Delta_{P}} MI_{Q}(X:Z|Y), \\
  \TSI(X:Y;Z) &= \max_{Q\in\Delta_{P}} CoI_{Q}(X;Y;Z), \\
  \TCI(X:Y;Z) &= MI(X:(Y,Z)) \\
  & \qquad\qquad- \min_{Q\in\Delta_{P}}MI_{Q}(X:(Y,Z)),
\end{align*}
where $MI$ denotes the mutual information, $CoI$ the coinformation (see Section~\ref{sec:mult-coinf} below), and the
index $Q$ in $MI_{Q}$ or $CoI_{Q}$ indicates that the corresponding information-theoretic quantity should be computed
with respect to the joint distribution~$Q$, as opposed to the ``true underlying distribution''~$P$.
As shown in~\cite{BROJA13:Quantifying_unique_information}, these four quantities are non-negative, and % they satisfy
\begin{equation}
  \label{eq:MI-decompositions}
  \begin{split}
    MI(X:(Y,Z)) & = \TSI(X:Y;Z) + \TUI(X:Y\setminus Z) \\
        & \quad + \TUI(X:Z\setminus Y) + \TCI(X:Y;Z), \\
    MI(X:Y) &= \TSI(X:Y;Z) + \TUI(X:Y\setminus Z), \\
    MI(X:Z) &= \TSI(X:Y;Z) + \TUI(X:Z\setminus Y).
  \end{split}
\end{equation}
Moreover, it was argued in~\cite{BROJA13:Quantifying_unique_information} that $\TSI(X:Y;Z)$ can be considered as a
measure of shared information, $\TCI(X:Y;Z)$ as a measure of complementary information, and $\TUI(X:Y\setminus Z)$ and
$\TUI(X:Z\setminus Y)$ as measures of unique information.  This interpretion can be justified by the following result,
which is a translation of some of the results of~\cite{BROJA13:Quantifying_unique_information}:
\begin{thm}
  Let $SI(X:Y;Z)$, $UI(X:Y\setminus Z)$, $UI(X:Z\setminus Y)$ and $CI(X:Y;Z)$ be non-negative functions on~$\Delta$
  satisfying an information decomposition of the form~\eqref{eq:MI-decompositions}, and assume that the following holds:
  \begin{enumerate}
  \item For any $P\in\Delta$, the maps $Q\mapsto UI_{Q}(X:Y\setminus Z)$ and $Q\mapsto
    UI_{Q}(X:Z\setminus Y)$ are constant on $\Delta_{P}$.
  \item For any $P\in\Delta$ there exists $Q\in\Delta_{P}$ with $CI_{Q}(X:Y;Z)=0$.
  \end{enumerate}
  Then $SI=\TSI$, $UI=\TUI$ and $CI=\TCI$ on~$\Delta$.
\end{thm}
Condition 1) says that the amount of unique information depends only
on the marginal distributions of the pairs $(X,Y)$ and~$(X,Z)$
formalizing the idea that unique information can be extracted from $Y$
and $Z$ alone independent of their joint distribution.  Condition
2) states that the presence or absence of synergistic information
cannot be decided from the marginal distributions alone.
See~\cite{BROJA13:Quantifying_unique_information} for a discussion of
these properties.

\smallskip %
In the present paper we ask how these results can be extended to the case of more variables.  The first question is how
the general structure of the decomposition should look like.  As stated above, a conceptional answer to this question is
given by the PI lattice of Williams and Beer.  However, as we will show in Section~\ref{sec:PI-vs-Id}, the
bivariate decomposition into the functions $\TSI$, $\TUI$ and $\TCI$ cannot be extended to this
framework.  The problem is that $\TSI$ satisfies the equality
\begin{equation*}
  \TSI((Y,Z):Y;Z) = MI(Y:Z), \tag{\emph{identity axiom}}
\end{equation*}
which was introduced in~\cite{HarderSalgePolani13:Bivariate_redundant_information}.  Theorem~\ref{thm:LP-vs-identity}
states that no non-negative information decomposition according to the PI lattice can satisfy the identity axiom.
Therefore, if there is a multivariate decomposition of $MI(X:(Y_{1},\dots,Y_{n}))$ that generalizes the information
decomposition into $\TSI$, $\TUI$ and $\TCI$ in a consistent way, then it cannot be a partial information decomposition.

% It is unclear how to extend these results to the case of more variables.  In the bivariate case it seems natural
% to look for a decomposition of the mutual information $MI(X:(Y,Z))$ into unique information, shared information and
% synergy, and many people would agree that there are only these three kinds of information (but of course, one can argue
% about this).  However, the situation is much less clear for more than two explanatory variables.  The \emph{partial
%   information decomposition} framework from~\cite{WilliamsBeer:Nonneg_Decomposition_of_Multiinformation} proposes a
% solution that we recall in Section~\ref{sec:PI-vs-Id}.  In this framework, $MI(X:(Y_{1},Y_{2},Y_{3}))$ is decomposed
% into 18 terms (Fig.~\ref{fig:PI3}).  However, as we will show in Theorem~\ref{thm:LP-vs-identity}, the bivariate
% decomposition into the functions $\TSI$, $\TUI$ and $\TCI$ cannot be extended to this
% framework, since $\TSI$ satisfies the \emph{identity axiom}
% \begin{equation*}
%   \TSI((Y,Z):Y;Z) = MI(Y:Z),
% \end{equation*}
% which was introduced in~\cite{HarderSalgePolani13:Bivariate_redundant_information}.  Therefore, if there is a
% multivariate decomposition of $MI(X:(Y_{1},\dots,Y_{n}))$ that generalizes the information decomposition into $\TSI$,
% $\TUI$ and $\TCI$ in a consistent way, then it cannot be a partial information decomposition.

Even without a consistent multivariate information decomposition the functions $\TSI$, $\TUI$ and $\TCI$ can be used in
the context of several variables by partitioning the variables.  For example, the quantity
\begin{equation*}
  \TUI(X:Y\setminus (Z_{1},\dots,Z_{n}))
\end{equation*}
should quantify the amount of information that only $Y$ knows about~$X$, but that none of the $Z_{i}$ has, and that also
none of the combinations of the $Z_{i}$ has.
In Section~\ref{sec:tsi-tui-tci-ineqs} we investigate what happens if we enlarge one of the arguments of the functions
$\TSI$, $\TUI$ and~$\TCI$.  In particular, we ask whether the functions increase or decrease in this case.

As shown in Section~\ref{sec:UI}, $\TUI$ behaves quite reasonable in this setting: $\TUI$ satisfies
\begin{align*}
  \TUI(X:Y \setminus(Z,Z')) &\leq \TUI(X:Y \setminus Z), \\
  \TUI(X:(Y,Y') \setminus Z) &\geq \TUI(X:Y \setminus Z), \\
  \TUI((X,X'):Y \setminus Z) &\geq \TUI(X:Y \setminus Z).
\end{align*}
Moreover, in Section~\ref{sec:SI} we show that $\TSI(X:(Y,Y');Z) \ge \TSI(X:Y;Z)$.  On the other hand, there is no
monotonic relation between $\TSI((X,X'):Y;Z)$ and $\TSI(X:Y;Z)$.  In particular, $\TSI$ does not satisfy the following
inequality, which was called \emph{left monotonicity} in~\cite{BROJ13:Shared_information}:
\begin{equation*}
  SI((X,X'):Y;Z) \ge SI(X:Y;Z).
\end{equation*}
Hence, enlarging $X$ may transform shared information into unique information.
Finally, in Section~\ref{sec:CI} we show that there is no monotonic relation between $\TCI(X:(Y,Y');Z)$
and~$\TCI(X:Y;Z)$, since the addition of $Y'$ may turn complementary information into shared information.  Moreover, there
is no monotonic relation between $\TCI((X,X'):Y;Z)$ and $\TCI(X:Y;Z)$ either.
% , that is, $\TCI$ does not satisfy a left monotonicity property.
Therefore, enlarging $X$ may transform complementary information into unique information.
We interprete our results in the concluding Section~\ref{sec:conclusions}.

\section{Mutual information and coinformation}
\label{sec:mult-coinf}

The mutual information is defined by
\begin{equation*}
  MI(X:Y) = H(X) + H(Y) - H(X,Y),
\end{equation*}
where $H(X) = - \sum_{x} p(X=x) \log p(X=x)$ denotes the Shannon entropy.  See~\cite{CoverThomas91:Elements_of_Information_Theory} for an interpretation
and further properties of~$MI$.
The mutual information satisfies the chain rule
\begin{equation*}
  MI(X:(Y,Z)) = MI(X:Y) + MI(X:Z|Y).
\end{equation*}
This identity can be derived from the entropy chain rule
\begin{equation*}
  H(X,Y) = H(Y) + H(X|Y).
\end{equation*}
Here, the conditional entropy and conditional mutual information are defined as follows: For any value $y$ of $Y$ with
$p(Y=y)>0$,
let $H(X|Y=y)$ and $MI(X:Z|Y=y)$ be the entropy and mutual information of random variables distributed according to the
conditional distributions $p(X=x|Y=y)$ and $p(X=x,Z=z|Y=y)$.  Then
\begin{align*}
  H(X|Y) &= \sum_{y}p(Y=y)H(X|Y=y) \\
\text{ and}\quad MI(X:Z|Y) &= \sum_{y}p(Y=y)MI(X:Z|Y=y).
\end{align*}

Chain rules are very important in information theory, and they also play an important role in the proofs in this paper.
Therefore, it would be nice if the quantities in an information decomposition would satisfy a chain rule.
Unfortunately, as discussed in~\cite{BROJ13:Shared_information}, this is not the case in any of the information
decompositions proposed so far.

The chain rule and non-negativity imply that $MI(X:(Y,Z)) \ge MI(X:Y)$.  This expresses the fact that ``more variables
contain more information.''

The coinformation of three random variables is defined as
\begin{equation*}
  CoI(X;Y;Z) = MI(X:Y) - MI(X:Y|Z).
\end{equation*}
Expanding $CoI(X;Y;Z)$ one sees that the coinformation is symmetric in its three arguments.  Moreover, the coinformation
satisfies the chain rule
\begin{equation*}
  CoI(X;(Y,Y');Z) = CoI(X;Y;Z) + CoI(X;Y';Z|Y).
\end{equation*}
However, since the coinformation is not non-negative, in general, it does not increase if one of the variables is
enlarged.

From~\eqref{eq:MI-decompositions} one can deduce
\begin{equation*}
  CoI(X;Y;Z) = \TSI(X:Y;Z) - \TCI(X:Y;Z).
\end{equation*}
This expresses the wellknown fact that a positive coinformation is a sign of redundancy, while a negative coinformation
indicates synergy.

\section{The partial information lattice and the identity axiom}
\label{sec:PI-vs-Id}

In this section we briefly recall the ideas behind the partial information (PI) lattice by Williams and Beer.  For
details we refer to~\cite{WilliamsBeer:Nonneg_Decomposition_of_Multiinformation}.  The PI lattice is a framework to
define information decompositions of arbitrarily many random variables.  Unfortunately, as we will show in
Theorem~\ref{thm:LP-vs-identity}, a non-negative decomposition of the mutual information according to the PI lattice is
not possible with the identity axiom.

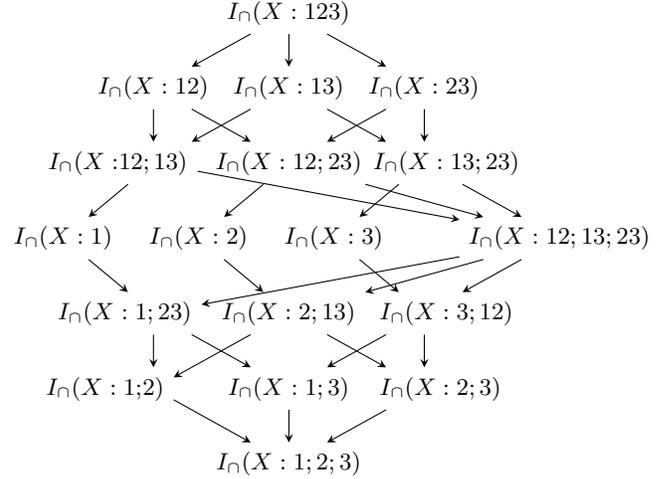
\begin{figure}
  \centering
    \begin{tikzpicture}[xscale=1.2]
    \PIdrei{\small $I_\cap(X:123)$}
           {\small $I_\cap(X:12)$}
           {\small $I_\cap(X:13)$}
           {\small $I_\cap(X:23)$}
           {\small $\mathllap{I_\cap(X:}12;13)$}
           {\small $I_\cap(X:12;23)$}
           {\small $I_\cap(X:13\mathrlap{;23)}$}
           {\small $I_\cap(X:1)$}
           {\small $I_\cap(X:2)$}
           {\small $I_\cap(X:3)$}
           {\small $I_\cap(X:12;13;23)$}
           {\small $\mathllap{I_\cap(X}:1;23)$}
           {\small $I_\cap(X:2;13)$}
           {\small $I_\cap(X:3\mathrlap{;12)}$}
           {\small $\mathllap{I_\cap(X:1;}2)$}
           {\small $I_\cap(X:1;3)$}
           {\small $I_\cap(X:2\mathrlap{;3)}$}
           {\small $I_\cap(X:1;2;3)$}
  \end{tikzpicture}
  \caption{The PI lattice for three variables.  For brevity, the sets $A_{i}$ are abbreviated by the indices of their
    elements; that is, $\{Y_{1},Y_{2}\}$ is abbreviated by $12$, and so on.}
  \label{fig:PI3}
\end{figure}

Consider $n+1$ variables $X, Y_{1},\dots,Y_{n}$.  We want to study in which way the information that $Y_{1},\dots,Y_{n}$
contain \emph{about~$X$} is distributed over the different combinations of the~$Y_{i}$.  For each subset
$A\subseteq\{Y_{1},\dots,Y_{n}\}$, the amount of information contained in $A$ is equal to the mutual
information~$MI(X:A)$ (where $A$ is interpreted as a random vector).  Different subsets
$A_{1},\dots,A_{k}\subseteq\{Y_{1},\dots,Y_{n}\}$ may share information, i.e.~they may carry redundant information.  % contain the ``same'', i.e. redundant or shared, information.
What we are looking for is a
function $I_{\cap}(X:A_{1};\dots;A_{k})$ to quantify this shared information.  Williams and Beer propose that this
function should satisfy the following axioms:
\begin{itemize}
\item $I_{\cap}(X:A_{1};\dots;A_{k})$ is symmetric under permutations of $A_{1},\dots,A_{k}$.
  \hfill\emph{(symmetry)}
\item $I_{\cap}(X:A_{1}) = MI(X:A)$.
  \hfill\emph{(self-redundancy)}
\item $I_{\cap}(X:A_{1};\dots;A_{k};A_{k+1}) \le I_{\cap}(X:A_{1};\dots;A_{k})$, with equality if $A_{i}\subseteq
  A_{k+1}$ for some~$i\le k$.
  \hfill\emph{(monotonicity)}
\end{itemize}
Any function $I_{\cap}(X:A_{1};\dots;A_{k})$ that satisfies these axioms is determined from its values on the
\emph{antichains}; that is, on the families $\{A_{1},\dots,A_{k}\}$ with $A_{i}\not\subseteq A_{j}$ for all~$i\neq j$.  The
antichains of subsets of $\{Y_{1},\dots,Y_{n}\}$ form a lattice with respect to the partial order
\begin{multline*}
  \{A_{1},\dots,A_{k}\} \preceq \{B_{1},\dots,B_{l}\} \\
  \Longleftrightarrow
  \text{ for each $B_{j}$ there is $A_{i}$ with $A_{i}\subseteq B_{j}$.}
\end{multline*}
This lattice is called the \emph{partial information (PI) lattice} in this context.  According to the
Williams-Beer-axioms, $I_{\cap}(X:\cdot)$ is a monotone function on this lattice.  The PI lattice for $n=3$ is depicted
in Fig.~\ref{fig:PI3}.

Let $A_{1},\dots,A_{k},A_{k+1}\subseteq\{Y_{1},\dots,Y_{n}\}$.
The idea behind the monotonicity axiom is, of course,
not only that the amount of redundant information in $A_{1},\dots,A_{k},A_{k+1}$ is less than the amount of redundant
information in $A_{1},\dots,A_{k}$ (when measured in bits), but that, in fact, the redundancy in
$A_{1},\dots,A_{k},A_{k+1}$ really is \emph{a part of} the redundancy in $A_{1},\dots,A_{k}$.  Similarly, in the case
that $A_{k}\subseteq A_{k+1}$, not only should the two amounts of redundant information agree, but they should really
refer to \emph{the same information}.  Therefore, in general, the difference
\begin{equation*}
  I_{\cap}(X:A_{1};\dots;A_{k}) - I_{\cap}(X:A_{1};\dots;A_{k};A_{k+1})
\end{equation*}
should measure the amount of information that is shared by $A_{1},\dots,A_{k}$, but that is not contained in~$A_{k+1}$.

Suppose that there exists a function $I_{\partial}(X:A_{1};\dots;A_{k})$ defined for any antichain
$\{A_{1},\dots,A_{k}\}$ that measure the amount of information contained in $I_{\cap}(X:A_{1};\dots;A_{k})$ that is not
contained in any of those terms $I_{\cap}(X:B_{1};\dots;B_{l})$ where the antichain
$\{B_{1},\dots,B_{l}\}\prec\{A_{1},\dots,A_{k}\}$.  
Then, if any information can be classified according to where, e.g. in which subset, 
it is available for the first time, e.g. it cannot be obtained from any smaller subset,
the following identity should hold:
\begin{equation*}
  I_{\cap}(X:A_{1};\dots;A_{k}) = \sum_{\makebox[0pt]{$\scriptstyle\{B_{1},\dots,B_{l}\}\preceq\{A_{1},\dots,A_{k}\}$}}I_{\partial}(X:A_{1};\dots;A_{k}).
\end{equation*}
As shown in~\cite{WilliamsBeer:Nonneg_Decomposition_of_Multiinformation}, this relation defines
$I_{\partial}(X:A_{1};\dots;A_{k})$ uniquely using the M\"obius inversion on the PI lattice.  In general, however, the
M\"obius inversion does not yield a non-negative function.
The property that $I_{\partial}$ is non-negative is called
\emph{local positivity} in~\cite{BROJ13:Shared_information}.  Using an idea from the same paper we now show that local
positivity contradicts the identity axiom mentioned in the introduction.

\begin{thm}
  \label{thm:LP-vs-identity}
  There are no functions $I_{\cap}$, $I_{\partial}$ that satisfy the Williams-Beer-axioms, local positivity and the
  identity axiom.
\end{thm}
\begin{IEEEproof}
  Suppose to the contrary that such functions do exist.
  Consider the case $n=3$, where $Y_{1},Y_{2}$ are independent
  uniformly distributed binary random variables, and where $Y_{3}=Y_{1}\XOR Y_{2}$.  Moreover, let
  $X=(Y_{1},Y_{2},Y_{3})$.  By the identity property, $I_{\cap}(\{Y_{i},Y_{j}\}:Y_{i};Y_{j})=MI(Y_{i}:Y_{j})=0\bit$ for any
  $i\neq j$.  Observe that any pair of the variables $\{Y_{1},Y_{2},Y_{3}\}$ determines the third random
  variable.  Therefore,   $X$ is just a relabeling of the state space $\{Y_i, Y_j\}$ for any
  $i\neq j$, and we obtain $I_{\cap}(X:Y_{i};Y_{j}) = I_{\cap}(\{Y_{i},Y_{j}\}:Y_{i};Y_{j}) = 0\bit$.
  By monotonicity, $I_{\cap}(X:Y_{1};Y_{2};Y_{3})=0\bit$, and so $I_{\cap}(X:\cdot)$ and
  $I_{\partial}(X:\cdot)$ vanish on the lower two levels of the PI lattice (Fig.~\ref{fig:PI3}).  On the next level, if
%  $(i,j,k)$ is a permutation of $(1,2,3)$, 
  $\{i,j,k\}=\{1,2,3\}$, then by identity $I_{\cap}(X:Y_{i};\{Y_{j}Y_{k}\}) =
  I_{\cap}(\{Y_{i},Y_{j},Y_{k}\}:Y_{i};\{Y_{j};Y_{k}\})=MI(Y_{i}:\{Y_{j},Y_{k}\})=1\bit$, and so
  $I_{\partial}(X:Y_{i};\{Y_{j},Y_{k}\}) = 1\bit$.  On the other hand,
  $I_{\cap}(X:\{Y_{1},Y_{2}\};\{Y_{1},Y_{3}\};\{Y_{2},Y_{3}\})\le MI(X:\{Y_{1},Y_{2},Y_{3}\})=2\bit$ by monotonicity, and
  so
  \begin{align*}
    I_{\partial}&(X:\{Y_{1},Y_{2}\};\{Y_{1},Y_{3}\};\{Y_{2},Y_{3}\}) \\
    & = I_{\cap}(X:\{Y_{1},Y_{2}\};\{Y_{1},Y_{3}\};\{Y_{2},Y_{3}\}) 
    \\ &\qquad - \sum_{\makebox[0pt]{$\scriptstyle\{i,j,k\}=\{1,2,3\}$}}I_{\partial}(X:Y_{i};\{Y_{j},Y_{k}\}) \\
    & \le 2\bit - 3\bit = -1\bit.
  \end{align*}
  This contradiction concludes the proof.
\end{IEEEproof}

\section[TUI, TSI and TCI in the multivariate setting]{$\TUI$, $\TSI$ and $\TCI$ in the multivariate setting}
\label{sec:tsi-tui-tci-ineqs}

% Notation: $SI$ for an arbitrary measure, $\TSI$ for our measure.
In this section we study what happens to the functions $\TUI$, $\TSI$ and $\TCI$ when one of their arguments is
enlarged.

\subsection{The unique information}
\label{sec:UI}

\begin{lemma}
  \begin{enumerate}
  \item $\TUI(X:Y \setminus(Z,Z')) \leq \TUI(X:Y \setminus Z)$.
  \item $\TUI(X:(Y,Y') \setminus Z) \geq \TUI(X:Y \setminus Z)$.
  \item $\TUI((X,X'):Y \setminus Z) \geq \TUI(X:Y \setminus Z)$.
  \end{enumerate}
\end{lemma}
\begin{IEEEproof}
  First we prove~1).  Let $P$ be the joint distribution of $X,Y,Z$, and let $P'$ be the joint distribution of $X,Y,Z,Z'$.  By
  definition, $P$ is a marginal of~$P'$.  Let $Q\in\Delta_{P}$, and let % define a distribution $Q'$ by
  \begin{multline*}
    Q'(x,y,z,z') :=
      \frac{Q(x,y,z)P'(x,z,z')}{P(x,z)}
      \\
      = Q(x,y,z) P'(z'|x,z)
  \end{multline*}
  if $P(x,z)>0$ and $Q'(x,y,z,z')=0$ else.  Then $Q'\in\Delta_{P'}$.  Moreover, $Q$ is the $(X,Y,Z)$-marginal of~$Q'$,
  and $Z'$ is independent of $Y$ given $X$ and~$Z$ with respect to~$Q'$.  Therefore,
  % \begin{equation*}
  %   MI_{Q}(X:Y|Z_{1},\dots,Z_{k}) = MI_{Q'}(X:Y|Z_{1},\dots,Z_{k})
  % \end{equation*}
  % and
  \begin{align*}
    MI_{Q'}&(X:Y|Z,Z') 
    \\ & = MI_{Q'}(X,Z':Y|Z) - MI_{Q'}(Z':Y|Z)
    \\ &
    \le MI_{Q'}(X,Z':Y|Z)
    \\ &
    = MI_{Q'}(X:Y|Z) + MI_{Q'}(Z':Y|X,Z)
    \\ &
    = MI_{Q'}(X:Y|Z) = MI_{Q}(X:Y|Z).
  \end{align*}
  The statement follows by taking the minimum over $Q\in\Delta_{P}$.

  Statements 2.~and 3.~can be proved together.  Consider five random variables $X,X',Y,Y',Z$ with joint
  distribution~$P'$, and let $P$ be the $(X,Y,Z)$-marginal of~$P'$.  Let $Q'\in\Delta_{P'}$, and let $Q$ be the
  $(X,Y,Z)$-marginal of~$Q'$.  Then $Q\in\Delta_{P}$.  Moreover,
  \begin{multline*}
    MI_{Q'}((X,X'):(Y,Y')|Z) \ge MI_{Q'}(X:Y|Z)
    \\ = MI_{Q}(X:Y|Z).
  \end{multline*}
  Taking the minimum for $Q'\in\Delta_{P'}$ implies
  \begin{equation*}
    \TUI((X,X'):(Y,Y') \setminus Z) \geq \TUI(X:Y \setminus Z).
  \end{equation*}
  Statements 2.~and 3.~follow by setting either $X'$ or $Y'$ to a constant random variable.
%
  % 2. Let $P'$ be the joint distribution of $X,Y,Y'$ and $Z$, and let $P$ be the $(X,Y,Z)$-marginal of~$P'$.  Let
  % $Q'\in\Delta_{P'}$, and let $Q$ be the $(X,Y,Z)$-marginal of~$Q'$.  Then $Q\in\Delta_{P}$.  Moreover,
  % \begin{equation*}
  %   MI_{Q'}(X:(Y,Y')|Z) \ge MI_{Q'}(X:Y|Z) = MI_{Q}(X:Y|Z).
  % \end{equation*}
  % The statement follows by taking the minimum for $Q'\in\Delta_{P'}$.
%
  % 3.
  % Let $P'$ be the joint distribution of $X,X',Y$ and $Z$, and let $P$ be the $(X,Y,Z)$-marginal of~$P'$.  Let
  % $Q'\in\Delta_{P'}$, and let $Q$ be the $(X,Y,Z)$-marginal of~$Q'$.  Then $Q\in\Delta_{P}$.  Moreover,
  % \begin{equation*}
  %   MI_{Q'}((X,X'):Y|Z) \ge MI_{Q'}(X:Y|Z) = MI_{Q}(X:Y|Z).
  % \end{equation*}
  % The statement follows by taking the minimum for $Q'\in\Delta_{P'}$.
\end{IEEEproof}

\subsection{The shared information}
\label{sec:SI}

\begin{lemma}
  $\TSI(X:(Y,Y');Z) \ge \TSI(X:Y;Z)$.
\end{lemma}
\begin{IEEEproof}
  Let $P'$ be the joint distribution of $X,Y,Y',Z$, and let $P$ be the $(X,Y,Z)$-marginals of~$P'$.
  For any $Q\in\Delta_{P}$ define a probability distribution $Q'$ by
  \begin{equation*}
    Q'(x,y,y',z) :=
    \begin{cases}
      \frac{Q(x,y,z)P'(x,y,y')}{P(x,y)}, & \text{ if }P(x,y)>0, \\
      0, & \text{ else.}
    \end{cases}
  \end{equation*}
  Then $Q'\in\Delta_{P'}$, and $Y'$ and $Z$ are conditionally independent given $X$ and $Y$ with respect to $Q'$.
  %
  % The chain rule of the coinformation states that
  % \begin{multline*}
  %   CoI_{Q'}(X,(Y,Y'),Z) \\ = CoI_{Q'}(X,Y,Z) + CoI_{Q'}(X,Y',Z|Y).
  % \end{multline*}
  Observe that $CoI_{Q'}(X,Y,Z) = CoI_{Q}(X,Y,Z)$ and
  \begin{multline*}
    CoI_{Q'}(X,Y',Z|Y) \\ = MI_{Q'}(Y',Z|Y) - MI_{Q'}(Y',Z|X,Y)
    \\ = MI_{Q'}(Y',Z|Y) \ge 0.
  \end{multline*}
  Hence, the chain rule of the coinformation implies that $CoI_{Q'}(X,(Y,Y'),Z) \ge CoI_{Q}(X,Y,Z)$.  The statement follows
  by maximizing~\mbox{$Q\in\Delta_{P}$}.
%  2. follows from 1. by symmetry.
\end{IEEEproof}

Should there be a relation between $SI((X,X'):Y;Z)$ and $SI(X:Y;Z)$?  In~\cite{BROJ13:Shared_information} the inequality
\begin{equation*}
  SI((X,X'):Y;Z) \ge SI(X:Y;Z)
\end{equation*}
is called \emph{left monotonicity}.  As observed in~\cite{BROJ13:Shared_information}, none of the
measures of shared information proposed so far satisfies left monotonicity.

$\TSI$ also violates left monotonicity.  Basically, the identity axiom makes it difficult to satisfy left monotonicity.
Consider two independent binary random variables $X,Y$ and let $Z=X\AND Y$.  Even though $X$ and $Y$ are independent,
one can argue that they share information about~$Z$.  For example, if $X$ and $Y$ are both zero, then both $X$ and $Y$
can deduce that~$Z=0$.  And indeed, in this example, $\TSI(Z:X;Y)\approx 0.311\bit$~\cite{BROJ13:Shared_information},
and also other proposed information decompositions yield a non-zero shared
information~\cite{HarderSalgePolani13:Bivariate_redundant_information}.  Therefore,
\begin{multline*}
  \TSI(Z:X;Y) > 0 = MI(X:Y)
  \\ = \TSI((X,Y):X;Y) = \TSI((Z,X,Y):X;Y).
\end{multline*}

As observed in~\cite{BROJ13:Shared_information}, a chain rule for the shared information of the form
\begin{equation*}
  SI((X,X'):Y;Z) = SI(X:Y;Z) + SI(X':Y;Z|X)
\end{equation*}
would imply left monotonicity.  Therefore, $\TSI$ does not satisfy a chain rule.

\subsection{The complementary information}
\label{sec:CI}

Should there be a relation between $CI(X:(Y,Y');Z)$ and $CI(X:Y;Z)$?  Since \emph{``more random variables contain more
  information,''} it is easy to find examples where \emph{more random variables contain more complementary
  information,''} that is $\TCI(X:(Y,Y');Z) > \TCI(X:Y;Z)$.  For example, let $Y'$, $Y$ and $Z$ be independent uniformly
distributed binary random variables and $X=Y'\XOR Z$.  In this example $Y$ and $Z$ know nothing about~$X$, but $Y'$ and
$Z$ together determine~$X$, and so
\begin{multline*}
  1\bit = \TCI(X:Y';Z) = \TCI(X:(Y,Y');Z) \\ > 0\bit = \TCI(X:Y;Z).
\end{multline*}

On the other hand, there are examples where $\TCI(X:(Y,Y');Z) < \TCI(X:Y;Z)$.  The reason is that further information
may transform synergistic information into redundant information.  For example, if $X=Y\XOR Z$, then
\begin{equation*}
  1\bit = \TCI(X:Y;Z) > 0\bit = \TCI(X:(Y,Z);Z).
\end{equation*}

Neither is there a simple relation between $\TCI((X,X'):Y;Z)$ and~$\TCI(X:Y;Z)$.  The argument is similar as for the
shared information. %~$\TSI$.
In fact, for any pair $(Y,Z)$ of random variables, the identity axiom implies $\TCI((Y,Z):Y;Z) =
0\bit$~\cite{BROJA13:Quantifying_unique_information}.
Consider again the case that $X=Y\XOR Z$.  As random variables, the triple $(X,Y,Z)$ is equivalent to the pair $(Y,Z)$.
Therefore,
\begin{multline*}
  1\bit = \TCI(X:Y;Z) > 0\bit \\ = \TCI((Y,Z):Y;Z) = \TCI((X,Y,Z):Y;Z).
\end{multline*}
So the left monotonicity for the synergy is violated again as a consequence of the identity axiom.  As above, this
implies that $\TCI$ does not satisfy a chain rule of the form
\begin{equation*}
  CI((X,X'):Y;Z) = CI(X:Y;Z) + CI(X':Y;Z|X).
\end{equation*}

\section{Conclusions}
\label{sec:conclusions}

We have seen that $\TUI$ behaves according to our intuition if one of its arguments is replaced by a ``larger random
variable.''  Moreover, $\TSI$ increases, if one of its right arguments is enlarged.  On the other hand, there is no
monotone relation for the left argument in $\TSI$, and for $\TCI$ there is no monotone relation at all.  In these last
cases, information is transformed in some way.  For example, if the inequality $\TCI(X:(Y,Y');Z) < \TCI(X:Y;Z)$ holds,
then the addition of~$Y'$ transforms synergistic information into redundant information. % as explained in~\ref{sec:CI}.

Let us look again at the example that demonstrates that $\TSI$ violates left monotonicity.  In the operational
interpretation of~\cite{BROJA13:Quantifying_unique_information} this has the following interpretation: If $Z=X\AND Y$,
then the two conditional distributions $p(Z=z|X=x)$ and $p(Z=z|Y=y)$ are identical.  Therefore, if $X$ or $Y$ can be used in a decision task which reward depends on ~$Z$, none of the two random variables performs better than the other; none of them has an advantage, and so
none of them has unique information about~$Z$.  On the other hand, $X$ and $Y$ do know different aspects about the
random vector $(X,Y)$, and depending on wether a reward function depends more on $X$ or on $Y$, 
they perform differently. Therefore, each of
them carries unique information about~$(X,Y)$. Intuitively, one could argue that combining the information in $X, Y$ 
via the $\AND$ function has transformed unique into shared information.

As stated above, the fact that $\TSI$ and $\TCI$ do not satisfy left monotonicity is related to the identity axiom.  For
the complementary information this relation is strict: Any measure of complementary information that comes from a
bivariate information decomposition of the form~\eqref{eq:MI-decompositions}, that satisfies the identity axiom and that
is positive in the $\XOR$-example violates left monotonicity, as the argument in Section~\ref{sec:CI} shows.  For the
shared information this relation is more subtle: Identity and left monotonicity do not directly contradict each other,
but whenever $X$ is a function of $Y$ and~$Z$ they imply the strong inequality $SI(X:Y;Z) \le MI(Y:Z)$.

In Section~\ref{sec:PI-vs-Id} we have shown that the identity axiom contradicts a non-negative decomposition according
to the PI lattice for~$n\ge 3$.  Therefore, if we want to extend the bivariate information decomposition into $\TUI$,
$\TSI$ and $\TCI$ to more variables, then this multivariate information decomposition must have a form that is different
from the PI lattice.  In particular, it is an open question which terms such an information decomposition should have.

Even if the structure of such a decomposition is presently unknown, we can interprete the bivariate quantities $\TUI$,
$\TSI$ and $\TCI$ in this context.  For example, the quantity $MI(X:Y_{1},\dots,Y_{k}) -
\sum_{i=1}^{k}\TUI(X:Y_{i}\setminus Y_{1},\dots,Y_{i-1},Y_{i+1},\dots,Y_{k})$ has the natural interpretation as
\emph{``the union of all information that is either synergistic or shared for some combination of variables.''}  Hence
we conjecture that this difference should be non-negative. 

The conjecture would follow from the inequality
\begin{multline*}
  \TUI(X:Y\setminus Z,W) + \TUI(X:Z\setminus Y,W) \\ \le \TUI(X:(Y,Z)\setminus W).
\end{multline*}
This inequality states that the unique information contained in a pair of variables is larger than the sum of the unique
informations of the single variables.  The difference between the right hand side and the left hand side should be due
to synergistic effects.  Proving (or disproving) the conjecture and this inequality would be a large step towards a
better understanding of the function~$\TUI$.

\subsection*{Acknowledgements}
JR acknowledges support by the VW foundation. NB acknowledges support by the Klaus Tschira Stiftung. EO has received funding
from the European Community's Seventh Framework Programme (FP7/2007-2013) under grant agreement no.~258749 (CEEDS) and
no.~318723 (MatheMACS).

\bibliographystyle{IEEEtran}
\bibliography{Blackwell}

\end{document}